\documentclass[twocolumn,amsmath,amssymb,prl]{revtex4}

\usepackage{graphicx}
\usepackage{dcolumn}
\usepackage{bm}

\begin{document}



\title{Test of cold denaturation mechanism for proteins as a function of water's structure
}

\author{Manuel I. Marqu\'es}
\email{manuel.marques@uam.es}

\affiliation{%
~Departamento de F\'isica de Materiales C-IV,
Universidad Aut\'onoma de Madrid, 28049 Madrid, Spain\\
}%



\begin{abstract}
 In a recent paper [PRL 91, 138103 (2003)] a new mechanism to explain the cold
denaturation of proteins, based on the loss of local low-density water structure, has been proposed.
In the present paper this mechanism is tested by means of full atom numerical simulations.
In good agreement with this proposal, cold denaturation resulting in the unfolded state was found at the High Density Liquid (HDL) state
of water, at which the amount of open tetragonal hydrogen bonds decreases at cooling.
 \end{abstract}


\maketitle



Hydrophobicity is an important driving force in the process of polymer folding and protein stabilization
 \cite{Privalov,Southall,Pratt,Kauzmann,Dill,DillII,Hummer,Chandler,Melchionna}.
Basically, hydrophobicity provides a mechanism to avoid the energy cost of disruption of a water
hydrogen bond by the presence of an apolar molecule .
Actually, the folding of a polymer is partially due to the assembly
of the apolar monomers in order to minimize the disruption
of hydrogen bonds among water molecules \cite{Kauzmann,Chandler,Stillinger}.
Also, the stability of the folded state is enhanced by the cage structures
formed of highly organized water molecules around the biopolymer where the disruption of hydrogen bonds is
minimized \cite{Frank}. Increasing the temperature of a polymer in a water bath leads to structures different from the
native folded state. Hydrophobicity is reduced by thermal fluctuations and the proteins stability
is lost. However it is also possible to reduce polymers hydrophobicity and to induce unfolded states by lowering
the temperature, a phenomenon called cold denaturation \cite{Pace,PrivalovII,Jonas}. The microscopic mechanism leading to
cancellation of the hydrophobic effects and cold denaturation is not clear, due in part to the complexity of
the protein-solvent interactions.
Cold denaturation has two important properties. (i) For pure water solvent, cold denaturation is only found
at very high pressures ($P>2kbar$) \cite{Kunugi}. So, in order to get high pressure cold destabilization of the folded protein,
hydrophobicity should decrease when
cooling at high pressures. (ii) The transition to the unfolded state is driven by a smaller volume, indicating a
positive slope in the P-T transition line. On the contrary, the usual thermal denaturation behaves with negative slope.
Hydrophobicity weakening at high pressures should also be accompanied by a positive slope.
In recent years there has been a huge amount of work in order the study the dependence
of hydrophobicity with pressure. Contact and solvent separated configurations have been studied from an enthalpy/entropy
point of view \cite{Ghosh}. Solvent separated configurations are increasingly stabilized at higher pressures by enthalpic contributions.
Full atom simulations of two Lennard-Jones particles in model water have found that the aggregation of the two particles becomes unstable
for pressures in the $kbar$ range \cite{Payne}, and recent replica exchange molecular dynamics simulations have studied the weakening
of the hydrophobic interaction with pressure for a C-terminal fragment of protein G \cite{Paschek0}.
These findings are in agreement with a cancellation of the
hydrophobic force at high pressures. Actually some recent molecular dynamics simulations have already found the swelling
of a protein when lowering the temperature at high pressures with a positive transition slope
as corresponding to the cold dentauration of real proteins \cite{Paschek}. However the question still remains: Why is
hydrophobicity canceled at very high pressures in such a way as to promote cold denaturation?

During the last years, several works have attempted to answer this question by using different models:
Thermodynamical models \cite{Hawley}, effective attraction models \cite{Vanderzande},
temperature dependent attraction models \cite{DillIII},effective water-protein
interactions models including solvent exclusion \cite{Hummer,Stillinger,PrattII}, lattice models \cite{Rios} based
on a bimodal description of the energy of water in the shell around the hydrophobic molecule \cite{Muller} and models
mimicking the interaction between water and non polar monomers \cite{Bruscolini}.
Very recently, a lattice model that captures the effect of volume correlations on water at high and low pressures
has been used to propose a possible mechanism for cold denaturation and hydrophobicity weakening at high pressures \cite{Marques}.
The proposal is based on the known existence of two possible water metastable states. A low density/pressure liquid with a
local structure very similar to the low-density amorphous (LDA) solid water, consisting of an open tetrahedral network,
and a high density/pressure water, very similar to high-density amorphous (HDA) solid water, where core-core repulsion
dominates, similar to simple, non anomalous liquids \cite{Starr}. A way to change from the low density liquid (LDL) to the high
density liquid (HDL) is by increasing the pressure at constant temperature. Actually, there is a critical pressure $P_{c}$ separating
both behaviors. For
$P<P_{c}$ a decrease in temperature implies an increase in the number of energetically favorable open tetrahedral interactions,
and an increase on volume. For a given temperature (called temperature of maximum density (TMD)) this increase in volume
is more important than the regular compresion of the liquid, and the density reaches a maximum. Then, the density decreases more and more
until reaching the negative sloped Ice Ih crystalline line ,or the negative sloped Kauzmann boundary (vanishing diffusivity) in numerical simulations.
 On the contrary
for $P>P_{c}$ the
number of interactions forming open tetrahedral structures decreases when cooling down the system, producing the more dense interpenetrating
 tetrahedral network and the more energetically
dominant core-core interactions. Now there is no TMD, and the density always decreases (similar to a non-anomalous liquid)
until reaching the positive sloped, Ice III or Ice V crystalline line (or the positive sloped Kauzmann boundary in numerical simulations).
Basically, for a constant temperature, the higher the pressure, the lower the number of low density open tetrahedral
structures. The hypothesis proposed in ref. \cite{Marques} is that hydrophobicity (i.e. aggregation to avoid the disruption of
hydrogen bonds and formation of open cage structures) is drastically reduced for the HDL state. Basically, at the HDL state, the existence of solvent-separated
apolar molecules is not energetically unfavorable as core-repulsion interaction is now the prevailing force instead of
the open tetrahedral hydrogen bonded network. Also, the stability due to the formation of low density hydrogen
bonded cage structures is not energetically favorable anymore. The high density water molecules are now capable of
penetrating inside the protein leading to unfolding \cite{Hummer}. With this hypothesis is easy to understand why cold dentauration
in pure water is found at $P>2kbar$, since $P_{c}\sim2kbar$ in pure water. Also it is possible to understand why the cold denaturation slope at the
P-T phase diagram is positive, since the non-anomalous, high pressure, ice crystalline lines (such as Ice III or Ice V) have also a positive slope (at $P>P_{c}$
the density of water always decreases when cooling down).
In order to check this idea, a simple lattice model was proposed in Ref. \cite{Marques} and the results were successfully compared with
the ones corresponding to real proteins such as Ribonuclease A \cite{Zhang}. However, a direct comprobation of this hypothesis by means of "real"
full atom simulations was still lacking. The aim of this paper is to present results from full atom simulations
which clearly correlate the weakening of hydrophobicity to the presence of the high density state of water.

In this paper the SPC/E model of water \cite{Berendsen} is simulated at different densities and temperatures. For positive pressures
the line of temperatures of maximum density (TMD) has a negative slope in the P-T plane and disappears for pressures
$P>P_c \sim 300MPa$ \cite{Scala}. From the point
of view of density-temperature simulations it implies that there is a critical density $\rho_{c} \sim 1.1 g/cm^{3}$.
If we do constant density simulations at $\rho < \rho_{c}$ we are at the LDL regime, and the number of open hydrogen bonded
tetrahedral networks grows as the temperature lowers. On the contrary, if we are at $\rho > \rho_{c}$ we are at the
HDL regime and the number of open hydrogen bonded tetrahedral networks decreases as lowering the temperature.
For this reason
, in order to study both regimes, we perform numerical simulations at $\rho=1g/cm^{3}, 1.1 g/cm^{3}, 1.2 g/cm^{3}$ and
$1.3 g/cm{3}$ (see Fig.1). At all densities, we perform simulations at $T=175K, 225K, 275K$ and $325K$ in order to sample from highly
thermally activated water molecules to conditions close to the Kauzmann boundary. First, in order to confirm the
correctness of the density-temperatures phase points chosen for the simulations, the
oxygen-oxygen pair correlation function for a sample with a maximum number of $460$ water molecules has been measured.
A reaction field technique \cite{Steinhauser} with a cut-off of $0.79nm$ accounts for the long-range Coulombic interactions. Configurations
were sampled by means of a non-constant step Monte Carlo algorithm.
Results are shown in Fig.2. They match previously reported calculations \cite{Starr} and recent experimental results \cite{Finney}.
At $\rho < \rho_{c}$ a maximum of the oxygen-oxygen correlation function at $r \sim 4.5 \AA$ is found. This maximum grows as we cool
down the system and is typical of LDA water and open tetrahedral network structures (like IceIh). It has been also measured
by numerical simulations in LDL water \cite{Starr}. On the contrary, at $\rho > \rho_{c}$, the peak at $r \sim 4.5 \AA$ decreases
when cooling down the system an a new peak at $r \sim 3.5 \AA$ rises up. This new peak is typical of the HDA water \cite{Finney} and
interpenetrating dense tetrahedral networks (like Ice VI and Ice VII). It has also been measured by computer simulations
in model water \cite{Starr}. With this checking we are now confident about the range of pressures to be simulated, and we know when
we are dealing with LDL and when we are dealing with HDL.

\begin{figure}
\centering
\includegraphics[width=7cm,height=7cm,angle=0]{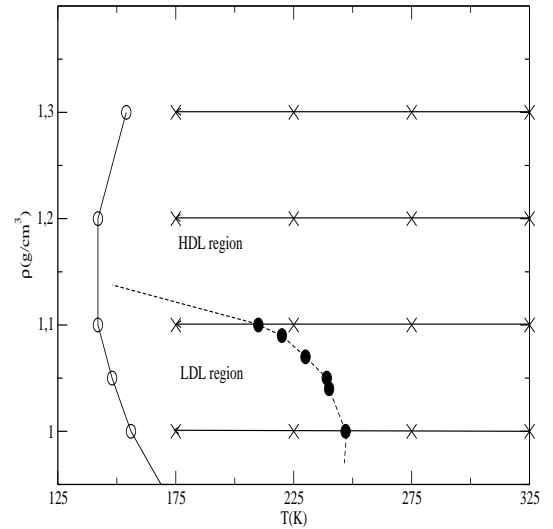}
\caption{Density-temperature phase diagram of SPC/E water. Black points correspond to the TMD line and white points correspond
the the Kauzmann boundary \cite{Scala}. Large crosses indicate the $\rho$ vs. $T$ conditions considered. Note how both
regimens (LDL and HDL) are taken into account.
}
\label{fig1}
\end{figure}

\begin{figure}
\centering
\includegraphics[width=7cm,height=7cm,angle=0]{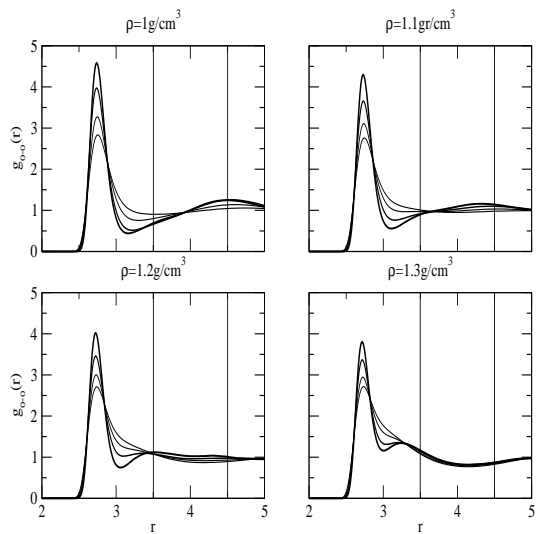}
\caption{Oxygen-Oxygen pair correlation functions vs. distance (in Angstroms) for the $\rho$ vs. $T$ conditions simulated in the paper. Thicker lines correspond
to lower temperatures. Vertical lines indicate the values of $r$ were main secondary peaks appear:
$r\sim3.5\AA$ for the HDL and $r\sim4.5\AA$ for the LDL.
}
\label{fig2}
\end{figure}

To test the hypothesis proposed in Ref. \cite{Marques} we should now embed an apolar polymer chain in our SPC/E water bath. If the hypothesis
is correct, at low temperatures, folded states are expected for $\rho < \rho_{c}$  and unfolded states are expected for
$\rho > \rho_{c}$. In order to attribute any configuration change to the influence of the solvent, we simulate a modified
model of the polyethylene molecule \cite{Larini}, free of bond angles and dihedral potential barriers. The polymer-chain consist of $N=15$ hydrophobic
$CH_{2}$ beads, represented by Lennard-Jones interaction sites with $\epsilon =0.112kcal/mol$ and $\sigma =1.33 \AA$ and linked to quasi-rigid harmonic bonds
of $1.53\AA$ length and interaction coupling $k_{r}=350kcal/mol$. The water-polymer cross parameters were obtained using the conventional
Lorentz-Berthelot mixing rules. The structure of the hydrophobic polyethylene chain is characterized by its
radius of gyration $R_{G}^{2}=1/N\sum((\vec{r_{i}}-\vec{r_{mc}})^{2})$ (being $\vec{r_{mc}}=1/N\sum\vec{r_{i}}$).
For the extended polymer
the radius of gyration is $R_{G}\sim 4\AA$ and a typical folded configuration with spherical-like symmetry has a gyration radius close to $R_{0}=2.75\AA$.
So, we consider a configuration to be folded
if the gyration radius is
$R_{G}<R_{0}=2.75\AA$ (see Fig.3).
For each pressure-
temperature point of the phase diagram we measured up to $10^{8}$ thermally equilibrated polymer configurations in order to
build a histogram with the most probable values for the gyration radius. Using these data we are able to calculate an averaged value
and the corresponding dispersion. Results are shown in Fig.4 (dispersions are represented as error bars).

\begin{figure}
\centering
\includegraphics[width=7cm,height=7cm,angle=-90]{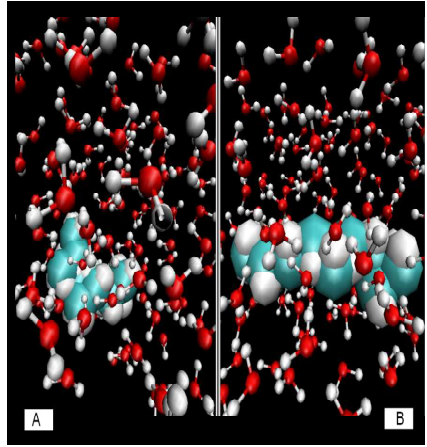}
\caption{(a) (Color online) Typical folded configuration of the polyethylene chain $R_{G}\sim2\AA$ and (b) typical unfolded configuration $R_{G}\sim3.5\AA$.
Atom's sizes are not at scale.}
\label{fig3}
\end{figure}

Clearly there are two different behaviors. For $\rho < \rho_{c}$ and low temperatures ($T\leq 275K$) the protein adopts a
folded state with a gyration radius $R_{G}<R_{0}$ (see Fig.4). There are no significant changes by cooling down
the system. The only way to obtain unfolded states is by increasing the temperature up to $T=325K$.
In this range of densities of the system, corresponding to the LDL state, no cold denaturation is found.
On the contrary when we increase the pressure up to $\rho > \rho_{c}$ (HDL state) there is a dramatic change on the
protein status by decreasing the temperature (see Fig 4). At low temperatures, the gyration radius grows to
a value corresponding to the unfolded state. We may definitively say that, in this range of densities of water
corresponding to HDL, cold denaturation is found at low temperatures.
In order to get a more direct relation between the value of the gyration radius and the water's tetragonal net status, it
is possible to define a parameter ($S=g_{o-o}(r=4.5\AA)-g_{o-o}(r=3.5\AA)/(g_{o-o}(r=4.5\AA)+g_{o-o}(r=3.5\AA))$) based on the
two secondary maxima of the oxygen-oxygen pair correlation function.
If the water's status corresponds to a LDL, the value of the correlation function at $r=4.5\AA$ is going to be much important
than the value of the correlation pair distribution function at $r=3.5\AA$ (see Fig.2), and $S$ is going to be positive.
If the status corresponds
to a HDL, the value at $r=3.5\AA$ is going to be larger than the value at $r=4.5\AA$ (see Fig.2)
and $S$ turns negative. Fig. 5 is a plot of the different gyration radius values found $R_{G}$ vs. $S$ (see Fig.5). The set gets divided in four quadrants:
(i) $R_{G}>R_{0}$ and $S>0$ (unfolded state and LDL water),(ii) $R_{G}>R_{0}$ and $S<0$ (unfolded state and HDL water),
(iii)$R_{G}<R_{0}$ and $S>0$ (folded state and LDL water), (iv) (1) $R_{G}<R_{0}$ and $S<0$ (folded state and HDL water).
(In order to avoid effects on the water structure due to thermal fluctuations we focus our attention on the low temperature regime $T<275K$).
 Note how almost all data corresponds to cases (ii) and (iii), in agreement with the hypothesis proposed in Ref. \cite{Marques}.

\begin{figure}
\centering
\includegraphics[width=7cm,height=7cm,angle=0]{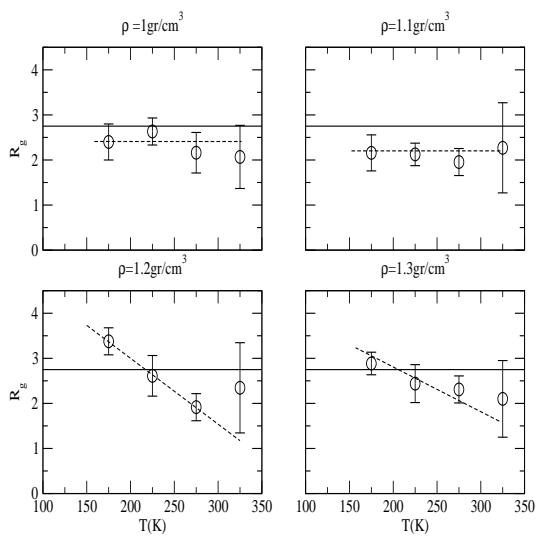}
\caption{Mean value and dispersion of the gyration radius (in Angstroms) for the polyethylene chain at the different $\rho$ vs. $T$ conditions considered
(dotted line is a guide to the eye).
Continuous line indicates the value $R_{0}=2.75\AA$.
At $\rho=1g/cm^{3}, 1.1 g/cm^{3}$ (LDL state) no cold denaturation is found by cooling and the gyration radius remains almost constant.
At $\rho=1.2g/cm^{3}, 1.3 g/cm^{3}$ (HDL state) cold denaturation is found and the gyration radius grows when cooling down.
}
\label{fig4}
\end{figure}

\begin{figure}
\centering
\includegraphics[width=7cm,height=7cm,angle=0]{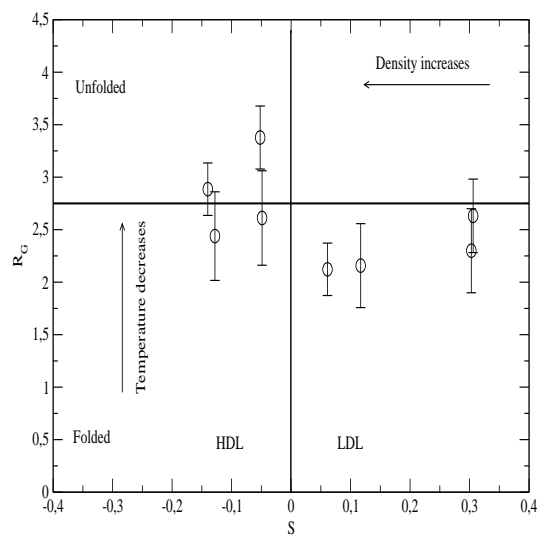}
\caption{Mean value and dispersion of the gyration radius (in Angstroms) vs. water's structure parameter ($S$) (see text).
Continuous horizontal line indicates a separation between folded
and unfolded states.
Continuous vertical line indicates the separation between the LDL and the HDL. Note how cold denaturation is mainly found at the HDL state.
}
\label{fig5}
\end{figure}

To conclude, full atom simulations of the folding behavior of a model polyethylene embedded in SPC/E water show
that there is a correlation between the cold denaturation and the density of water. For LDL the existence of a local structure
consisting of an open tetrahedral hydrogen bonded network prevents cold denaturation. The volume of this network increases
when cooling down the system and so does hydrophobicity. On the contrary, for the high pressure HDL, the local structure
is a packed tetrahedral hydrogen bonded network dominated by core-core repulsions. The volume of this network decreases when
cooling down the system and so does hydrophobicity, explaining the existence of high pressure cold denaturation of proteins.

Comments from J.A.Gonzalo and H.E.Stanley are greatfully acknowledged.

\end{document}